# Improving photon number resolvability of a superconducting nanowire detector array using a level comparator circuit

Jia Huang[1, 2†], Xingyu Zhang[1, 2†], Weijun Zhang[1, 2, 3*], Chaomeng Ding[1, 2], Yong Wang[1, 2], Chaolin Lv[1, 2], Guangzhao Xu[1, 2, 3], Xiaoyu Liu[1, 2], Hao Li[1, 2, 3], Zhen Wang[1, 2, 3], and Lixing You[1, 2, 3, 4*]

[1]National Key Laboratory of Materials for Integrated Circuits, Shanghai Institute of Microsystem and Information Technology (SIMIT), Chinese Academy of Sciences (CAS), 865 Changning Road., Shanghai, 200050, China

[2]CAS Center for Excellence in Superconducting Electronics, 865 Changning Road, Shanghai, 200050, China

[3]Center of Materials Science and Optoelectronics Engineering, University of Chinese Academy of Sciences, 19A Yuquan Road, Beijing, 100049, China

[4]State Key Laboratory of Marine Resource Utilization in South China Sea, Hainan University, 58 Renmin Avenue, Haikou, Hainan, 570228, China

[†]These authors contributed equally: Jia Huang, Xingyu Zhang

[*]Correspondence authors. Email: zhangweijun@mail.sim.ac.cn and lxyou@mail.sim.ac.cn

**Abstract**

Photon number resolving (PNR) capability is very important in many optical applications, including quantum information processing, fluorescence detection, and few-photon-level ranging and imaging. Superconducting nanowire single-photon detectors (SNSPDs) with a multipixel interleaved architecture give the array an excellent spatial PNR capability. However, the signal-to-noise ratio (SNR) of the photon number resolution ($SNR_{PNR}$) of the array will be degraded with increasing the element number due to the electronic noise in the readout circuit, which limits the PNR resolution as well as the maximum PNR number. In this study, a 16-element interleaved SNSPD array was fabricated, and the PNR capability of the array was investigated and analyzed. By introducing a level comparator circuit (LCC), the $SNR_{PNR}$ of the detector array was improved over a factor of four. In addition, we performed a statistical analysis of the photon number on this SNSPD array with LCC, showing that the LCC method effectively enhances the PNR resolution. Besides, the system timing jitter of the detector was reduced from 90 ps to 72 ps due to the improved electrical SNR.

Keywords: photon number resolving, superconducting nanowire single-photon detector, signal-to-noise ratio

## 1. Introduction

Superconducting nanowire single-photon detectors (SNSPDs) are excellent detectors for performing experiments in various fields, such as quantum optics [1], quantum communication [2], life science [3], and deep-space laser communication [4], owing to their high system detection efficiency [5], low dark count rate (DCR) [6], high count rate (CR) [7], and excellent time resolution [8, 9]. Besides these features, the photon number resolving (PNR) capability is a key parameter in quantum information processing that takes advantage of multiphoton states, and it would be particularly useful in linear optical quantum computing [10], quantum key distribution [11], quantum repeaters [12] and nonclassical state generation and detection [13].

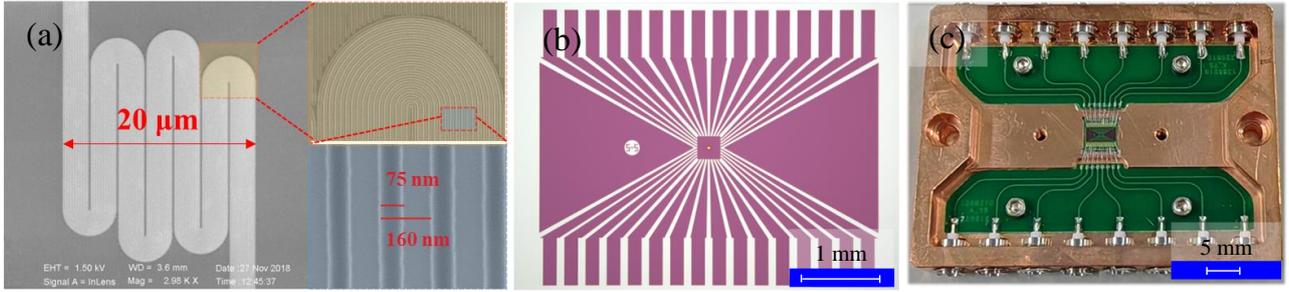

**Fig. 1** (a) SEM image of a 16-interleaved nanowire SNSPD array, with an active area of 20 μm in diameter. Insets of (a): zoomed-in SEM images of the rounded bends and the nanowires respectively. The bends are designed to reduce the current crowding effect. The nanowires are 75 nm wide and 160 nm in pitch. (b) Optical photograph of an SNSPD array chip with a 16-pixel electrode layout. The regions covered by NbN film are shown in purple. (c) Photograph of a chip-mounting block used for the 16-pixel SNSPD array with the fiber holder uninstalled.

However, the output of a conventional SNSPD, which is designed as a single meandered nanowire, is not proportional to the number of photons in the detector, and thus, PNR capability cannot be achieved using a conventional circuit. To overcome this challenge, alternative approaches to construct a PNR-SNSPD include temporal-multiplexing [14], spatial-multiplexing [15, 16], hybrid spatiotemporal-multiplexing [17], impedance-matched SNSPDs [18]. An interleaved multiplexing SNSPD array is one of the mainstream spatial-multiplexing SNSPDs due to its excellent comprehensive performance. In 2009, an interleaved PNR-SNSPD array was first proposed by Dauler et al [15]: interleaved nanowires were formed into a 4-pixel SNSPD array, the nanowires shared the same active area and each nanowire was biased and read out individually. A quasi-PNR capacity could be demonstrated by combining the $n$-port readout electronics, owing to the position-resolving capability of photons hitting different nanowires simultaneously. The voltage pulses from the fired different nanowires can be added up, producing a pulse amplitude proportional to the number of photons. So far, the maximum resolved photon number reported for an interleaved SNSPD array is 16 [7, 19].

The PNR fidelity (probability of correctly measuring $n$ incident photons) of the interleaved SNSPD array is mainly limited by the SDE and the number of pixels in the array [15]. However, the PNR resolution (pulse discrimination) is mainly limited by the signal-to-noise ratio (SNR) of PNR ($SNR_{PNR}$). For example, in the process of acquiring $n$-photon response pulse waveform by an oscilloscope, the difference in amplitude of neighboring pulses decreases as the number of response photons increases, due to the limitation of $SNR_{PNR}$. Specifically, $SNR_{PNR}$ of such detectors is mainly attributed to two following factors: first is the electrical noise of the readout circuit, which is the primary [7, 20]; second is the non-uniformity of the photon-response pulses of the nanowires in different channels, such as the slight differences in amplitudes, time jitters, and time delays, resulting in a further decay in the PNR capability after the pulse combination [21]. Thus, $SNR_{PNR}$ also partially limited the maximum PNR number (dynamic range) of the detector array.

In this study, to improve PNR resolution, a level comparator circuit (LCC) was added to a conventional readout circuit. By including the LCC, $SNR_{PNR}$ was improved by more than a factor of four. In addition, owing to the improved electrical SNR of the output pulse, the system timing jitter of a single pixel was reduced from ~90 ps to ~72 ps. This approach provides a simple, straightforward, and easily scalable approach to external circuit processing, making it attractive for use in future array-based PNR single photon detectors.

## 2. Device design, fabrication, and packaging

Figure **1** (a) shows a scanning electron microscope (SEM) image of the fabricated device. The device consisted of 16 interleaved nanowires that formed a circular active area with a diameter of 20 μm. The interleaved design ensured that all nanowires were illuminated equally, thus improving the PNR capability compared with a spatial array consisting of a series of multiple nanowires [15]. The bends of each nanowire were rounded to minimize the current crowding effect [22, 23].

The device was fabricated on a 2-inch Si substrate that was thermally oxidized on both sides. A multilayer anti-reflection coating consisting of five alternating layers of $SiO_2/TiO_2$ was deposited on the back of the substrate to enhance the transmittance to the substrate [24]. An NbN film with a thickness of 6.5 nm was then deposited on the front of the substrate by reactive DC magnetron sputtering in an Ar/N$_2$ gas mixture under a total pressure of 0.27 Pa. The Ar and N$_2$ flow rates were set to 30 and 4 sccm, respectively, using gas-flow controllers. Next, the NbN film was patterned using electron beam lithography and reactive ion etching to produce nanowires that were 75-nm wide and that had a 160-nm pitch. A quarter-wave optical stack consisting of a SiO layer with a thickness of 210 nm and a metallic mirror (comprising a 5-nm thickness of Ti and 100-nm thickness of Au) was deposited on the nanowires to enhance photon absorption at 1550 nm. The electrode pads, which are shown in Fig. **1** (b), were fabricated using ultraviolet lithography and reactive ion etching.

After fabrication, the SNSPD array was wire-bonded to the pads of a printed circuit board and soldered to small p-type connectors. The chip was then aligned and packaged with a single-mode lens fiber using backside illumination at room temperature. The chip-mounting block was mounted on a 2-K stage of a closed-cycle G-M cryocooler, as shown in Fig. 1 (c) with the fiber holder removed. Finally, the array was cooled down to 2.2-K operating temperature for further measurements.

### 3. Experiment design and setup

Figure **2** shows a typical histogram of the pulse-height distribution for an *n*-photon response. For quantitative analysis, the $SNR_{PNR}$ of a PNR-SNSPD is defined as $SNR_{PNR} = (H(n) − H(n − 1))/V_n = \Delta H/V_n$ [25], where *n* is the number of fired nanowires, $H(n)$ and $V_n$ represent the peak and the full width at half maximum (FWHM) values of the Gaussian distribution of the *n*-photon output amplitude, respectively.

Since the $SNR_{PNR}$ directly depends on the ratio between $\Delta H$ and $V_n$ of the pulse amplitude distribution, we can improve $SNR_{PNR}$ either by increasing $\Delta H$ through the enhanced gain of the readout circuit or decreasing $V_n$ through the reduced fluctuations in waveform amplitude. After investigation, we found that an LCC was a suitable device to meet the aforementioned requirements, which contains a specialized high-gain differential amplifier and a voltage-level converter, outputting a digital signal with high and constant amplitude. Thus, with a proper LCC design, we can increase the gain of the circuit and reduce the influence of fluctuations in waveform amplitude simultaneously, i.e., improve the $SNR_{PNR}$.

Figure **3** (a) shows the schematic diagram of our homemade LCC, comprising a voltage comparator (Analog Devices Inc., AD96685BR), a level translator (ON Semiconductor, MC100EPT25), power supply modules (supplying power to the voltage comparator and the level translator), and an adjustable resistor. The working process of LCC is as follows: First, an input pulse signal ($V_{in}$) is transmitted to the input port of the voltage comparator. If $V_{in}$ is higher than a reference trigger voltage ($V_{tr}$), the voltage comparator converts the input analog signal into a digital square-wave signal ($V_s$) with constant amplitude. The output pulse width is determined by the width between the rising and falling edges of the input waveform at the selected trigger level $V_{tr}$. The value of $V_{tr}$ can be tuned by adjusting the adjustable resistor, satisfying the condition $0 < V_{tr} < V_{in}$. Then, $V_s$ is further processed by the level translator and translated to a higher-level output voltage ($V_{out}$).

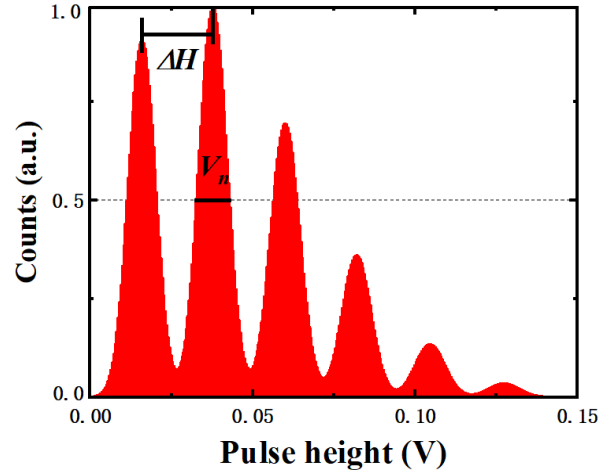

**Fig. 2** Typical histogram of the pulse height distribution for an interleaved SNSPD, the mean photon number is $\mu = 5$.

Figure **3** (b) shows a self-developed 10-channel LCC board, which includes four sections: the core components (consisting of the voltage comparators, the level translators, and the adjustable resistors), the "power supply" section (which supplies power to the core components), the "input signal port" section and the "output signal port" section.

Figure **3** (c) illustrates the original LCC input signal, and Fig. **3** (d) shows the output signal by applying LCC. The LCC improved the $SNR_{PNR}$ in two aspects: first, in the process of the level conversion into a digital square wave, the large fluctuation of the amplitude in the input pulse signal was significantly suppressed, because the converted pulse was triggered at the fast, low-jitter rising edge of the input pulse, and the information on the amplitude was omitted in this process; second, the low-level output from the voltage comparator was amplified to a higher-level signal by the level translator.

The schematic setup of an interleaved PNR-SNSPD array combined with an LCC readout circuit is shown in Fig. **4** (a). The voltage pulse generated from each nanowire of the SNSPD array is amplified using a room-temperature 50 dB low-noise amplifier (LNA650, RF Bay Inc.). Unlike the previous readout circuit [18], the signal from each nanowire is first connected to a single channel LCC and then output through a power combiner. The use of multi-channel LCC also helps to eliminate the non-uniformity of the photon-response pulses between channels. We used a pulsed laser with a repetition rate of 10 MHz and a wavelength of 1550 nm, the photon number distribution in the laser pulse corresponds to Poisson statistics and the input photon flux emitted by the pulsed laser could be controlled using serial variable attenuators.

Selecting an appropriate $V_{tr}$ value is crucial for minimizing the timing jitter ($T_j$) of the readout circuits. We

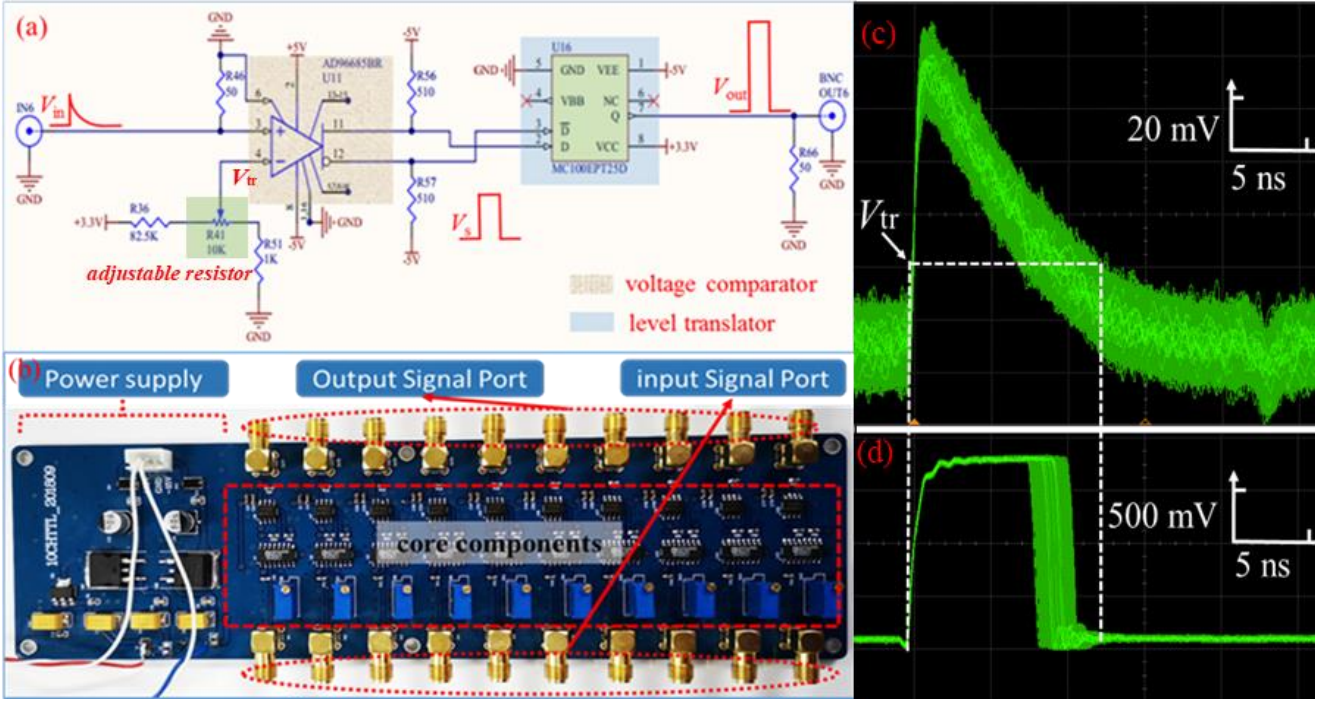

**Fig. 3** (a) Schematic diagram of the self-developed LCC. (b) Photo of a 10-channel LCC board. The board is divided into four major sections: power supply, input signal port, output signal port, and the core components (comprising of the voltage comparators, level translators, and adjustable resistors). (c) Oscilloscope persistence map of the input signal (20 mV/div). (d) Oscilloscope persistence map of the corresponding output signal by applying LCC (500 mV/div).

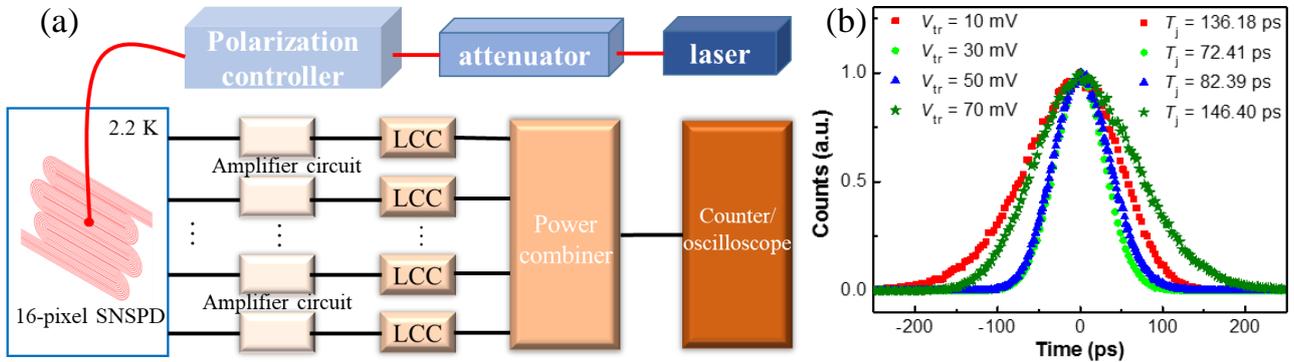

**Fig. 4** (a) Schematic setup of an interleaved PNR-SNSPD array with LCC integrated circuit. (b) $T_j$ of a single pixel SNSPD determined at different $V_{tr}$ of LCC, which is the full width at half maximum (FWHM) of the histogram.

investigated $T_j$ of the LCC integrated circuit for different values of $V_{tr}$, as shown in Fig. 4 (b). It is found that the smallest $T_j$ was obtained for $V_{tr} = 30$ mV. Larger $T_j$ values were obtained for $V_{tr} = 10$ mV and $V_{tr} = 70$ mV because the former is close to the electrical noise floor, and the latter is close to the amplitudes of the fluctuations in pulse. There is a slight difference between $T_j$ values for $V_{tr} = 50$ mV and 30 mV, mainly due to the difference in the slope of the rising edge of the pulse signal at the discrimination level. As a result, a value of $V_{tr} = 30$ mV was finally chosen for the subsequent experimental measurements.

Figure 5 (a) shows a comparison of the output waveforms of a single pixel under different readout circuits with and without LCC. Further, we collected the distribution statistics of the pulse height corresponding to these two cases, as shown in the inset of Figure 5 (a). The distribution statistics of the pulse height follow a Gaussian distribution. We then defined a quantity to describe the SNR of the pulse waveform, denoted as $SNR_{signal} = H/N$, where $H$ and $N$ represent the peak and the

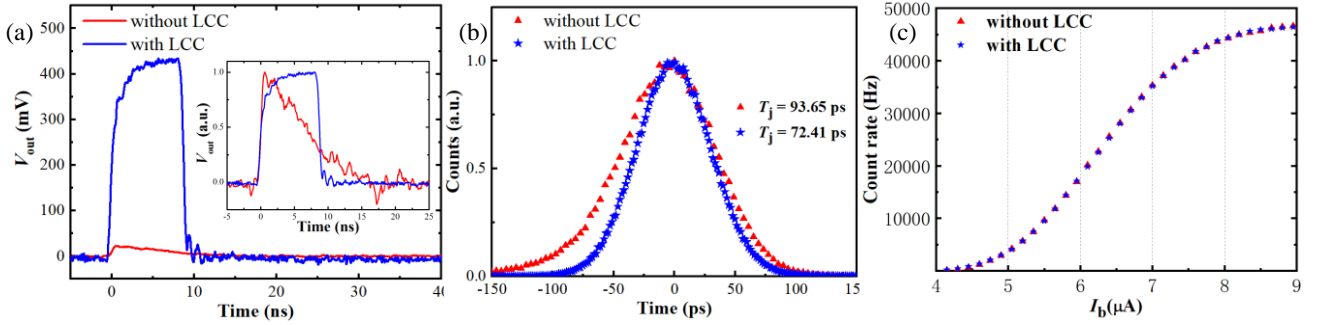

**Fig. 5** (a) Output waveforms of a single pixel recorded at different readout circuits with (blue line) and without (red line) LCC. The inset represents the statistics of the corresponding pulse height. (b) $T_j$ of a single pixel recorded at different readout circuits with and without LCC. The trigger level of LCC was set to 30 mV. (c) Count rate comparisons for circuits with and without LCC, the mean photon number is $\mu = 0.1$. The trigger level of LCC was set to 30 mV.

FWHM values of the Gaussian distribution of signal amplitude fluctuation, respectively. It was found that, with LCC, $H$ increased significantly from ~20 mV to ~420 mV, while $N$ changed from 3.5 mV to 25 mV; thus, the $SNR_{signal}$ calculated by its definition was improved from 5.7 to 16.8, with an enhanced factor of ~2.9. Figure **5** (b) shows the comparison between $T_j$ values for circuits with and without LCC. It can be seen that the circuit with the LCC has a significantly lower $T_j$ due to the improved $SNR_{signal}$ of the waveform, this is thus another important advantage of using the LCC. Figure **5** (c) shows the comparison between the count rate for circuits with and without LCC. It can be seen that the count rates are basically the same with and without LCC, indicating that the introduction of LCC does not influence the device count rates.

### 4. Results and discussion

We then characterized the PNR capability and $SNR_{PNR}$ of our interleaved SNSPD array for different readout setups. Figs. **6** (a) and (b) show the persistence maps recorded by an oscilloscope under different readout setups without and with LCC, respectively. The histogram represents the statistics of pulse height. It can be seen that with LCC included, the individual voltage peaks, which represent different numbers of detected photons, are clearly separated from each other and the pulse width of the peaks is small, indicating that the photon numbers are well resolved.

Figures **6** (c) and (d) show the plots of $SNR_{PNR}$ against $n$ for readout circuits without and with the LCC, respectively, recorded at two values of the mean photon number $\mu$. For $\mu = 10$ and $n = 8$, the $SNR_{PNR}$ values without LCC are ~2 ($\Delta H \approx$ 20 mV, $V_n \approx 10$ mV), whereas those with LCC are ~8 ($\Delta H \approx$ 420 mV, $V_n \approx 50$ mV). Consequently, by applying the LCC, the $SNR_{PNR}$ values increased by a factor of ~3.8 on average (4.1 in maximum).

Notably, in the comparisons of the circuits with and without LCC, the enhanced factor of $SNR_{PNR}$ (~3.8, for multichannel) is larger than that of $SNR_{signal}$ (~2.9, for a single channel in Figure 5(a)). This is because the contributions of $SNR_{PNR}$ of multiple channels include the $SNR_{signal}$ of a single channel and the pulse non-uniformity between multiple channels. In the process of $n$-photon detection, the superposition of pulse signals from multiple channels is improved by the LCC circuit, i.e., the pulse non-uniformity between channels is reduced, resulting in improved $SNR_{PNR}$. For example, the ratio of $V_n$ to $N$ in readout circuits without LCC is ~2.9, while that in readout circuits with LCC is 2. A smaller value of $V_n$ to $N$ indicated a better pulse consistency in the measurement.

We then performed the photon count statistics of the PNR capability using a counter. Since the counter in our laboratory could only read signals with amplitudes in the range from −300 mV to 300 mV, we connected a −29 dB electrical attenuator in series with the power combiner to reduce the output amplitude from ~420 to ~16 mV, which consequently ensures that all of the signals were detected by the counter when all 16 pixels responded simultaneously. Although the transmitted signal and noise are both strongly attenuated by the same attenuator, due to the additional noise of the back-end measurement equipment, the final measured noise amplitude $V_n$ value is ~2.2 mV; thus, the calculated $SNR_{PNR}$ value is ~7 (slightly lower than that without the attenuator (~8)). We measured the number of counts in a certain time interval (1 s in this study) for different counter threshold voltages and for different mean photon numbers; the results obtained are shown in Fig. **7** (a). The existence of 16 flat regions demonstrates the operation of all pixels and 16 photon levels of the detector: The flat region 1 corresponds to the detection of at least one (≥1), region 2 of ≥ 2, region 3 of ≥ 3, …, 14 (≥ 14), region 15 of ≥ 15, and finally region 16 of 16 photons. Consequently, the photon count rate for the PNR detector can be conveniently measured with a pulse counter by triggering the corresponding flat region and subtracting the count rate of the next higher amplitude region. As shown in Fig. **7** (b), the CR was measured as a function of $\mu$ by setting different trigger levels of the counter, with DCR subtracted. We chose the value corresponding to the center of the flat region in order from 1 to 16 shown in Fig. 7 (a) as the threshold voltage for the absorption of ≥1, ≥2, ≥3, …, ≥14, ≥15, ≥16 photons. According to [23], when the mean photon number is low so that $\eta\mu \ll 1$ ($\eta$ is the device quantum efficiency obtained with respect to $\mu$), the CR should be approximately proportional to the value of $(\eta\mu)^n$. This is confirmed by the results shown in Fig. 7 (b); the solid lines are the linear fits to the data with slopes very close to 1, 2, 3, …, and 7, which confirms that the detector responds to 1–7 photon detection events.

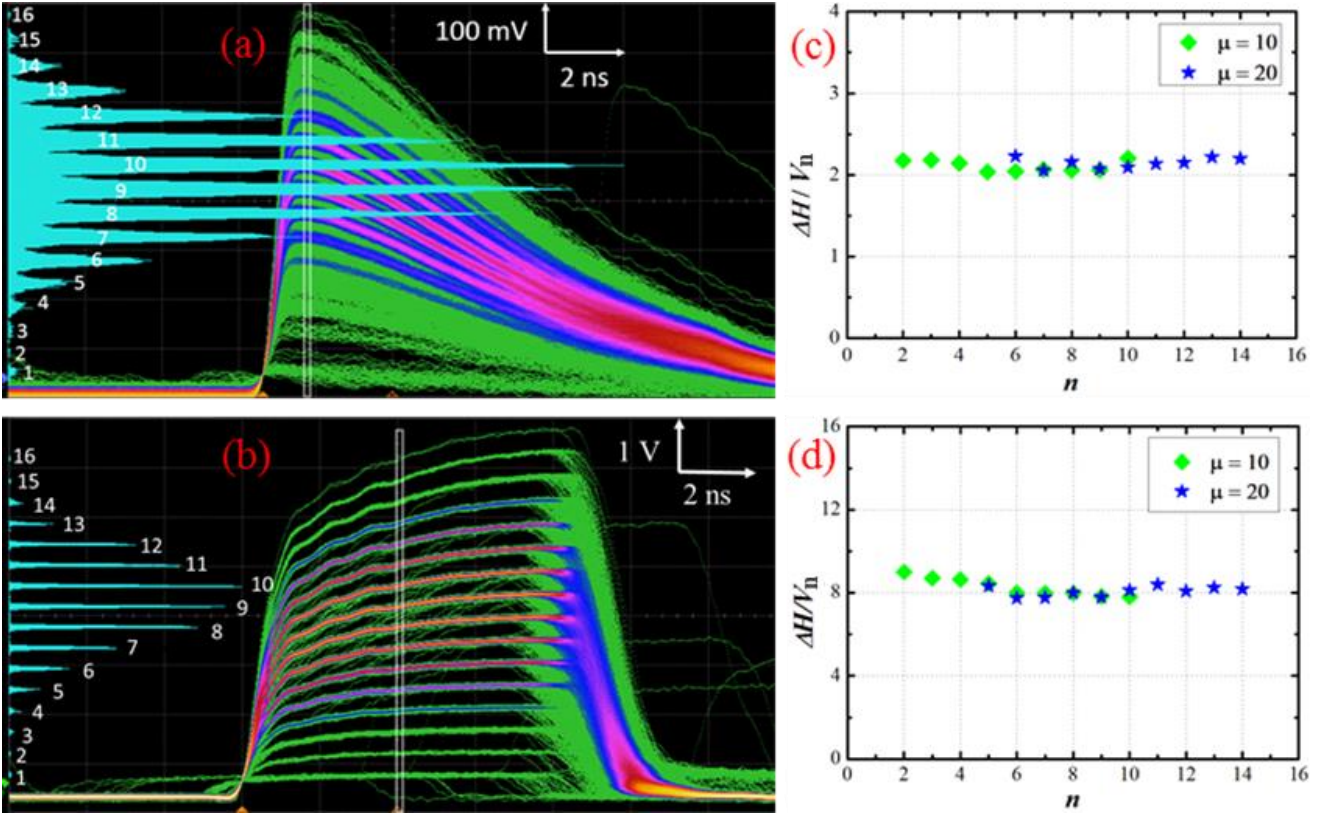

**Fig. 6** PNR capability and $SNR_{PNR}$ for different readout setups. Oscilloscope persistence maps of the PNR response: (a) without LCC; (b) with LCC. The histogram of the y-axis represents the distribution statistics of pulse height. The mean photon number is $\mu = 20$. Plots of $\Delta H/V_n$ vs the number of fired nanowires $n$ for readout setups at different mean photon numbers $\mu$: (c) without LCC; (d) with LCC.

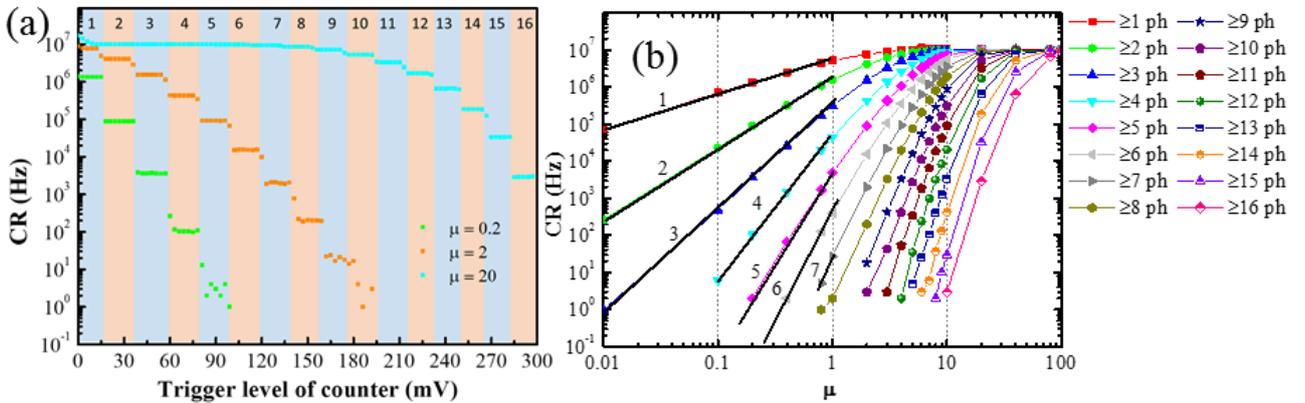

**Fig. 7** Detector array with an LCC integrated circuit: (a) Count-rate dependence on the counter threshold voltage for different mean numbers of $\mu$. (b) CR plotted as a function of $\mu$, corresponding to the detections of "≥n photons" (1≤n≤16).

## 5. Conclusions

We proposed and verified an LCC readout scheme for improving the $SNR_{PNR}$ of the multi-element PNR-SNSPD pulses. Compared to the original circuit, the $SNR_{PNR}$ is improved by a factor of ~4. Besides, the system timing jitter for a single pixel was reduced from ~90 ps to ~72 ps owing to the improved electrical SNR. The proposed readout circuit operates at room temperature and can be implemented easily. We expect the proposed multi-element PNR-SNSPD system will find applications that require PNR with high resolution.

**Acknowledgments**

This work is sponsored by the Shanghai Sailing Program (Grants No. 21YF1455700 and 21YF1455500), the National Natural Science Foundation of China (NSFC, Grants No. 12033007 and 61971409), and the Youth Innovation Promotion Association of the Chinese Academy of Sciences (Grants No. 2019238 and 2021230).